\documentclass[12pt]{article}
\usepackage{epsfig}

\newcommand{\beq}{\begin{equation}}
\newcommand{\eeq}{\end{equation}}
\newcommand{\beqa}{\begin{eqnarray}}
\newcommand{\eeqa}{\end{eqnarray}}

\def\noj#1,#2,{{\bf #1} (19#2)\ }
\def\jou#1,#2,#3,{{\sl #1\/ }{\bf #2} (19#3)\ }
\def\ann#1,#2,{{\sl Ann.\ Physics\/ }{\bf #1} (19#2)\ }
\def\cmp#1,#2,{{\sl Comm.\ Math.\ Phys.\/ }{\bf #1} (19#2)\ }
\def\ma#1,#2,{{\sl Math.\ Ann.\/ }{\bf #1} (19#2)\ }
\def\ng#1,#2,{{\sl Nagoya.\ Math.\ J.\/ }{\bf #1} (19#2)\ }
\def\jd#1,#2,{{\sl J.\ Diff.\ Geom.\/ }{\bf #1} (19#2)\ }
\def\invm#1,#2,{{\sl Invent.\ Math.\/ }{\bf #1} (19#2)\ }
\def\cq#1,#2,{{\sl Class.\ Quantum Grav.\/ }{\bf #1} (19#2)\ }
\def\cqg#1,#2,{{\sl Class.\ Quantum Grav.\/ }{\bf #1} (19#2)\ }
\def\ijmp#1,#2,{{\sl Int.\ J.\ Mod.\ Phys.\/ }{\bf A#1} (19#2)\ }
\def\jmphy#1,#2,{{\sl J.\ Geom.\ Phys.\/ }{\bf #1} (19#2)\ }
\def\jams#1,#2,{{\sl J.\ Amer.\ Math.\ Soc.\/ }{\bf #1} (19#2)\ }
\def\grg#1,#2,{{\sl Gen.\ Rel.\ Grav.\/ }{\bf #1} (19#2)\ }
\def\mpl#1,#2,{{\sl Mod.\ Phys.\ Lett.\/ }{\bf A#1} (19#2)\ }
\def\nc#1,#2,{{\sl Nuovo Cim.\/ }{\bf #1} (19#2)\ }
\def\np#1,#2,{{\sl Nucl.\ Phys.\/ }{\bf B#1} (19#2)\ }
\def\pl#1,#2,{{\sl Phys.\ Lett.\/ }{\bf #1B} (19#2)\ }
\def\pla#1,#2,{{\sl Phys.\ Lett.\/ }{\bf #1A} (19#2)\ }
\def\pr#1,#2,{{\sl Phys.\ Rev.\/ }{\bf #1} (19#2)\ }
\def\prd#1,#2,{{\sl Phys.\ Rev.\/ }{\bf D#1} (19#2)\ }
\def\prl#1,#2,{{\sl Phys.\ Rev.\ Lett.\/ }{\bf #1} (19#2)\ }
\def\prp#1,#2,{{\sl Phys.\ Rept.\/ }{\bf #1C} (19#2)\ }
\def\ptp#1,#2,{{\sl Prog.\ Theor.\ Phys.\/ }{\bf #1} (19#2)\ }
\def\ptpsup#1,#2,{{\sl Prog.\ Theor.\ Phys.\/ Suppl.\/ }{\bf #1} (19#2)\ }
\def\rmp#1,#2,{{\sl Rev.\ Mod.\ Phys.\/ }{\bf #1} (19#2)\ }
\def\yadfiz#1,#2,#3[#4,#5]{{\sl Yad.\ Fiz.\/ }{\bf #1} (19#2) #3%
\ [{\sl Sov.\ J.\ Nucl.\ Phys.\/ }{\bf #4} (19#2) #5]}
\def\zh#1,#2,#3[#4,#5]{{\sl Zh.\ Exp.\ Theor.\ Fiz.\/ }{\bf #1} (19#2) #3%
\ [{\sl Sov.\ Phys.\ JETP\/ }{\bf #4} (19#2) #5]}

\begin{document}
\vspace*{-.6in}
\thispagestyle{empty}
\begin{flushright}
IASSNS-HEP-99/29\\
hep-th/9903125
\end{flushright}
\baselineskip = 20pt

\vspace{.5in}
{\Large
\begin{center}
\textbf{Brane mapping under $(-1)^{F_L}$}
\end{center}}

\begin{center}
Keshav Dasgupta\footnote{E-mail:  keshav@sns.ias.edu}, 
Jaemo Park\footnote{E-mail:  jaemo@sns.ias.edu} \\
\emph{School of Natural Sciences, Institute for Advanced Study \\ 
Princeton, NJ 08540 , USA}
\end{center}

\vspace{1in}

\begin{center}
\textbf{Abstract}
\end{center}
\begin{quotation}
\noindent We present the evidence that $(-1)^{F_L}$ is a nonperturbative 
symmetry of Type II string theory. We argue that $(-1)^{F_L}$ is a 
symmetry of string theory as much as the $SL(2, Z)$ of the Type II 
string is and how the branes are mapped under the $(-1)^{F_L}$ modding. 
NS branes are mapped into the NS branes with the same world volume 
dimensions but with the different chiral structure. 
Supersymmetric Dp-branes (bound with anti-Dp branes)
are mapped to unstable nonsupersymmetric 
Dp-branes, which has D(p$-1$)-branes as kink solutions according to 
Horava\cite{Ho}.

\end{quotation}
\vfil

\newpage

\pagenumbering{arabic}

 In duality of string theory, it is known that S-duality and orbifolding 
do not always commute\cite{Sen, VaWi}. 
One example is the relation between the 
worldsheet orientation reversal $\Omega$ and $(-1)^{F_L}$ which flips 
the sign of the left-moving Ramond sector. In many of the situations 
$\Omega$ and $(-1)^{F_L}$ are S-dual operations\cite{VaWi}. 
Since Type IIB string 
theory is self-dual, we expect that by modding out by $\Omega$ and 
$(-1)^{F_L}$ respectively, we obtain the new dual pair. 
However, this does not hold. Type IIB modded by $\Omega$ gives rise 
to Type I string theory while type IIB modded by $(-1)^{F_L}$ is the 
Type IIA string theory. 
There are two important differences in the above example. 
Type IIB string modded by $(-1)^{F_L}$ is a typical example of the 
orbifold of the closed string theory while Type I string theory 
is not an orbifold of the closed string theory. 
Instead of the twisted sector, in Type I string theory 
D-branes are  introduced for the 
cancellation of the tadpole. Thus the non-perturbative states 
are introduced for the consistency of the theory but the rule 
of introducing the suitable D-branes are quite different from that 
obtaining the twisted sectors of the orbifold\cite{GiPo}. 
A related fact is that we obtain the other half of the gravitini with 
different chirality in the twisted sector in $(-1)^{F_L}$ orbifold, 
while the half of the gravitini are projected out 
by the presence of the D-branes 
in Type I theory. 

The recent paper by Hull and the subsequent papers\cite{Hull, Hull2} 
shed some light on this puzzle. 
He gives some evidence for the existence of NS-9 branes and argues that 
we can obtain SO(32) heterotic string if we consider the Type IIB 
modded by $(-1)^{F_L}$ {\it in the background of 32 NS-9 branes}. 
Since this is the string theory in the background of non-perturbative 
states and we do not have the sufficient understanding of 
the conformal field theory of NS branes, we cannot immediately 
prove or rule out this claim. Also the existence of NS-9 branes are 
still problematic, since there are no charges of the perturbative 
Type II string to couple to NS-9 branes. 
But some of the arguments are plausible, at least if we assume the 
existence of NS-9 branes. In analogy with the Type I theory as a string 
theory with the nonperturbative background, we can imagine that we 
should not introduce the twisted sector as in the orbifold case.
Also the presence of the NS-9 brane breaks the half of the supersymmetry 
if the NS-9 branes is a supersymmetric object forming a $SL(2, Z)$ 
doublet with D-9 brane. In his paper\cite{Hull}, Hull raises 
one interesting issue. He claims that $(-1)^{F_L}$ is a full 
non-perturbative symmetry of the Type IIB in the presence of 
NS-9 branes and he shows how $(-1)^{F_L}$ acts on non-perturbative 
states. This begs the question: Can we define $(-1)^{F_L}$ as a 
non-perturbative symmetry of Type II string theory {\it without} 
the presence of NS-9 branes? The purpose of this paper is to show 
some affirmative evidences for this question. In particular we argue that 
$(-1)^{F_L}$ is a nonperturbative symmetry as much as $SL(2,Z)$ is 
one of Type IIB theory. We shall see how $(-1)^{F_L}$ acts on NS branes 
and D-branes of Type II theory. Thus we give some evidence that 
Type IIB modded out by $(-1)^{F_L}$ is a Type IIA string theory 
in a full nonperturbative sense.  

In M-theory, if $(-1)^{F_L}$ corresponds to a full non-perturbative 
symmetry, then $(-1)^{F_L}$ corresponds to the operation 
$X^{11} \rightarrow -X^{11}$ and $A \rightarrow -A$ where $X^{11}$ is 
the eleventh-circle whose size is related to the coupling constant of 
Type IIA string theory and $A$ is the 3-form of the 11-d supergravity. 
The combined operation of $(-1)^{F_L}$ and the orientation reversal 
$R_9\Omega$ which acts as 
$X^{9} \rightarrow -X^{9}$ and $A \rightarrow -A$   
is mapped to a generator of $SL(2, Z)$ of Type IIB theory 
under the correspondence of M-theory on $T^2$ and Type IIB on a circle. 
Thus the claim that $(-1)^{F_L}$ is a full nonperturbative symmetry 
is equivalent to the self-duality of Type IIB-theory together with the 
consistency of the perturbative string theory. 
If $(-1)^{F_L}$ is not a full nonperturbative symmetry, 
then all facts of string dualities are in danger, which is clearly 
implausible. 

Since we know how $(-1)^{F_L}$ acts on perturbative states, it is 
interesting to see how $(-1)^{F_L}$ acts on nonperturbative states 
of Type II theory. Let us first start with NS-5 branes. For this 
purpose we can recall the fact that type IIA(IIB) theory on 
$A_{k-1}$ singularities can be mapped to type IIB(IIA) 
$k$ coincident NS-5 branes via T-duality\cite{Wi}. 
If we consider $k$ Kaluza-Klein monopole configurations in Type IIA(IIB) 
side, the transverse geometry is described by $(R^3\times S^1)/Z_k$
in the coincidence limit. The configuration has $U(1)$ isometry and 
we can T-dualize along the $U(1)$-direction, which gives us NS-5 
branes in Type IIB(IIA) side. By the explicit calculation at the 
orbifold point $T^4/Z_2$ of K3, we can see that Type IIA theory on K3 
is mapped 
to Type IIB on K3 under $(-1)^{F_L}$. Since $A_{k-1}$ singular point  
can be obtained by a suitable noncompact limit of K3, we can conclude 
that Type IIA(IIB) NS-5 branes are mapped to Type IIB(IIA) NS-5 
branes. From this consideration, we see that the chirality of NS-5 
branes are related to that of Type II string theory. Probably the 
conformal field theory describing the NS-5 brane has the different 
GSO projection in NSR formalism, which leads to different chiral 
structure. This is true of the conformal field theory describing 
the near horizon geometry of NS-5 brane 
in Type II theory\cite{CHS, ABS}. The worldsheet fermions of the 
conformal field theory are free and different GSO projections for 
left-moving and right-moving fermions lead to the different chiral 
structure of Type IIA and Type IIB NS-5 branes in the near horizon 
limit. Same thing happens for the NS-1 brane, i.e., the 
fundamental string and we expect the similar for NS-9 branes.

Now we turn into D-branes of Type II string theory. 
If we consider a D-p brane whose worldvolume spans $x^0, x^1, \cdots, 
x^p$, the suitable boundary condition at the worldsheet 
for the bosonic coordinates 
are given by 
\begin{equation}
\partial_n X^{\mu}=0 \,\,\, \mu=0, \cdots, p  \label{eq:bc}
\end{equation}
\begin{equation}
X^{\mu}=constant \,\,\, \mu=p+1, \cdots, 9 \nonumber
\end{equation}  
where $\partial_n$ is the derivative normal to the boundary of the 
worldsheet. 
This should be supplemented by a suitable boundary condition 
for the fermionic coordinates. Since the boundary condition 
(\ref{eq:bc})  
does 
not change under $(-1)^{F_L}$, it seems that a Dp-brane of Type IIA(IIB) 
is mapped to a Dp-brane of Type IIB(IIA). An immediate problem is that 
Type IIB string theory has supersymmetric D-branes with even dimension 
of worldvolume while 
Type IIA string theory has supersymmetric ones with odd dimension 
of worldvolume. Thus it's not 
clear how to obtain supersymmetric D-branes of Type IIA(IIB) starting 
from Type IIB(IIA) supersymmetric D-branes. One clue comes from a 
recent paper by Horava\cite{Ho}. He showed that 
in Type IIA side, all stable 
D-brane configuration can be constructed as a bound state 
of unstable Type IIA 
D9-branes. However at this point we note that its not natural to study
the action of $(-1)^{F_L}$ on a {\it single} D9 brane as under this
 operation it will naturally transform to anti-D9 brane. Therefore we
will study the action of $(-1)^{F_L}$ on the GSO invariant boundary
state and from there create brane configurations which are invariant
under this operation.  
  Thus we will show that a Type IIB D9 brane- anti D9 brane is mapped to 
the Type IIA D9 brane. From here we can recover all 
stable D-brane configurations in the type IIA theory. To start,  
we represent a D9-brane as a boundary state using open-closed string 
duality\cite{CaPo}. 
Since the boundary state is made up of the closed string 
perturbative states, we can easily figure out how $(-1)^{F_L}$  acts 
on the boundary state. 

The boundary state corresponding to D9-brane in the light cone gauge 
following the convention of \cite{Bergman} is given by, 
\begin{equation}
|B, \eta>^{NSNS}_{RR}=exp( \sum_{\mu=2}^9(-\sum_{n=1}^{\infty} 
\frac{1}{n}\alpha^{\mu}_{-n}\tilde{\alpha}^{\mu}_{-n}
-\sum_{r > 0}i\eta\psi^{\mu}_{-r}\tilde{\psi}^{\mu}_{-r}))
|B, \eta>^{(0)NSNS}_{RR}
\end{equation}
where $\eta=\pm$, $n$ is an integer and $r$ is a half-integer for NSNS 
sector and an integer for RR sector. The oscillator without tilde 
represents the right mover and the one with tilde represents the 
left mover.   
$|B, \eta>^{(0)}$ denotes the Fock vacuum 
with zero momentum state. This  specifies the unique state for NSNS 
sector. RR sector is more subtle due to the fermionic zero modes, which 
we will discuss shortly.   
The boundary state is invariant  under the type IIB GSO projection
\begin{equation}
\frac{1}{4}(1+(-1)^{F})(1+(-1)^{\tilde{F}}).
\end{equation}
where $(-1)^{F}$ and $(-1)^{\tilde{F}}$ flips the sign of 
$\psi^{\mu}_{-r}$ and $\tilde{\psi}^{\mu}_{-r}$.  
We take $|\eta>^{(0)NSNS}\equiv |0>^{NSNS}$ to be odd under
$(-1)^{F}$ and $(-1)^{\tilde{F}}$. Thus the action of $(-1)^{F}$ and 
$(-1)^{\tilde{F}}$ is given by
\begin{equation}
(-1)^{F}|B, \eta>^{NSNS}=(-1)^{\tilde{F}}|B, \eta>^{NSNS}
=-|B, -\eta>^{NSNS}.
\label{eq:nsns}
\end{equation} 
The GSO invariant configuration is 
\begin{equation}
|B, +>^{NSNS}-|B, ->^{NSNS}.
\end{equation}

The GSO operation $(-1)^{F}$ and $(-1)^{\tilde{F}}$ act on the left 
and right Ramond 
ground state as
\begin{equation}
(-1)^{F}\equiv \prod \sqrt{2}\psi^{\mu}_0, \,\, 
(-1)^{\tilde{F}}\equiv \prod \sqrt{2}\tilde{\psi}^{\mu}_0. \label{eq:gso}
\end{equation}
respectively where $\psi^{\mu}_0$ and $\tilde{\psi}^{\mu}_0$ are 
fermionic zero modes. 
We define the left and right 
Ramond ground state is even under $(-1)^{F}$ and  
$(-1)^{\tilde{F}}$ respectively. Finally we define $|B, \eta>$ to be a 
RR ground state satisfying 
\begin{equation}
(\psi^{\mu}_0+i\eta \tilde{\psi}^{\mu}_0)|B, \eta>^{(0)}_{RR}=0.
\end{equation}
We choose the relative normalization between $|B, +>$ and $|B, ->$ 
by setting
\begin{equation}
|B, +>^{(0)}_{RR}=\prod \frac{1}{\sqrt{2}}(\psi^{\mu}_0
+i\tilde{\psi}^{\mu}_0)|B, ->^{(0)}_{RR}.
\end{equation}
>From this and the anticommuation relation $\{\psi^{\mu}_0, \psi^{\nu}_0\}
=\delta^{\mu\nu}$ and the same for $\tilde{\psi}^{\mu}_0$ and 
$\tilde{\psi}^{\nu}_0$, we can show that  
\begin{equation}
|B, +>^{(0)}_{RR}=(-1)^{F}|B, ->^{(0)}_{RR}=(-1)^{\tilde{F}}
|B, ->^{(0)}_{RR}  \label{eq:rr}
\end{equation}
and 
\begin{equation}  
(-1)^{F}|B,\eta>_{RR}=(-1)^{\tilde{F}}|B,\eta>_{RR}=|B, -\eta>_{RR}.
\end{equation}  
Thus the GSO invariant state is given by
\begin{equation}
|B +>^{RR}+|B ->^{RR}.
\end{equation}
Note that the sign of $(-1)^F$ in (\ref{eq:nsns}) and 
(\ref{eq:rr}) is the same as $(-1)^{\tilde{F}}$ 
for the boundary state and the resulting boundary state survives 
Type IIB GSO projection. 

Now if we take the $(-1)^{F_L}$,  NSNS boundary state is invariant 
while RR part is projected out since the sign of the D-brane charges are 
flipped under $(-1)^{F_L}$ and the sign of the RR charge is related to 
the relative sign between NSNS boundary state and RR boundary 
state\cite{Sen2}. Thus in the untwisted sector 
only NSNS part survives. The twisted boundary condition 
is the same for NSNS sector, but the GSO projection 
for the RR sector is changed into 
\begin{equation}
\frac{1}{4}(1+(-1)^{F})(1-(-1)^{\tilde{F}}).
\end{equation}
By construction, the sign of $(-1)^{F}$ is the same as that of 
$(-1)^{\tilde{F}}$ for $|B+>_{RR}$ and $|B->_{RR}$. Thus 
no linear combination
of these survive the GSO projection in the twisted sector. 
Thus we just have NSNS boundary state under $(-1)^{F_L}$, which is 
exactly the Type IIA D9-brane state of Horava. Now in the original 
IIB theory this NSNS boundary state is nothing but a bound state of
D9, anti D9 system. Therefore we see that a D9, anti D9 system of type
IIB theory goes to an unstable D9 brane of type IIA\footnote{We would 
like to
thank Ashoke Sen for pointing this out to us. See also Ref\cite{senii}
where related results are derived using different methods.}. 
By similar calculation using the boundary state 
corresponding to Dp brane, one can show that Type IIB Dp, 
anti Dp brane system  
for odd p is mapped to Type IIA Dp brane under $(-1)^{F_L}$. 
If Type IIA Dp brane is unstable, Type IIA supersymmetric D-(p-1) brane 
emerges as a kink solution of Dp brane\cite{Ho}.

{\bf Acknowlegment}
 We thank  
 E. Witten for useful discussions. The work of K.D and J.P is supported 
in part by the U.S. Dept. of Energy under Grant No.
DE-FG02-90-ER40542.

\end{document}